\documentclass[superscriptaddress,preprint]{revtex4-1}
\usepackage{graphicx}
\usepackage{amsmath}
\usepackage{multirow}

\date{\today}

\begin{document}

\title{Large-scale vorticity generation and kinetic energy budget along the U.S. West Coast}

\author{G\'abor T\'oth}
\affiliation{Department of Physics of Complex Systems, ELTE  E\"otv\"os Lor\'and University, Budapest, Hungary.}
\author{Vikt\'oria Homonnai},
\affiliation{Department of Physics of Complex Systems, ELTE  E\"otv\"os Lor\'and University, Budapest, Hungary.}
\affiliation{Hungarian Meteorological Service, Budapest, Hungary.}
\author{Imre M. J\'anosi}
\affiliation{Department of Physics of Complex Systems, ELTE  E\"otv\"os Lor\'and University, Budapest, Hungary.}



\begin{abstract}
We attempt to evaluate energy budget over
a restricted but extremely well studied oceanic region along the shorelines of Oregon and California. The analysis is
based on a recently updated geostrophic flow field data set covering 22 years with daily resolution on a grid of 
0.25$^\circ\times$0.25$^\circ$, and turbulent wind stress data from the ERA-Interim reanalysis over the same geographic
region with the same temporal and spatial resolutions. Integrated 2D kinetic energy, enstrophy, wind stress work and {  kinetic
energy tendency} are determined separately for the shore- and open water regions. 
The empirical analysis is supported by 2D lattice Boltzmann simulations of freely decaying vortices along a rough solid wall,
which permits to separate the pure shoreline effects and dissipation properties of surface flow fields. 
Comparisons clearly demonstrate that
kinetic energy and vorticity {  of the geostrophic flow field} are mostly generated along the shorelines and advected to the open water regions, where
the net wind stress work is almost negligible.  Our results support that the geostrophic flow field is quasistationary on the
timescale of a couple of days, thus total forcing is practically equal to total dissipation. Estimates of unknown 
terms in the equation of oceanic kinetic energy budget are based on other studies, nevertheless our results
suggest that {  an effective} eddy kinematic viscosity is in the order of magnitude $10^{-2}$ m$^2$/s along the shorelines,
and it is  lower  {  by a factor of two} in the open water region. 
\end{abstract}

\maketitle

\section{Introduction}

A recent effort by \citet{Risien15} extends the work of  \citet{Saraceno08} by improving 
the temporal resolution from weekly to daily sea level anomaly fields, over a longer period, 
and by expanding the region to include the entire U.S. West Coast. The result
is a geostrophic flow field resolving {  mesoscale} eddies in details. The original
data set is exploited by studying transport of coastal zooplankton communities in the same
region \citep{alga11}, the use of high-frequency radar coastal currents to correct satellite altimetry
\citep{Roesler13},
 the abundance of two subarctic
oceanic copepods  in relation to regional ocean conditions \citep{alga15}, or the annual cycle of sea level in coastal areas from gridded satellite 
 altimetry and tide gauges \citep{gauge15}.
 
 Here we evaluate this surface flow data supplemented with ERA Interim  {  wind
 stress} time series, and attempt to perform a kinetic energy budget analysis over the
 studied geographic area. 
The domain is divided into two areas, one is close to the shoreline
 and another one is representing open water. The separated evaluation reveals
 the distinguished role of the interaction of oceanic flows with the solid boundaries
 in shallow water: most of the kinetic energy and vorticity {  of the geostrophic flow field}, thus  mesoscale
  vortices are generated in the coastal region. The findings are supported
 by 2D Lattice Boltzmann simulations, which help to identify the most essential
 physical components of the surface flow and the effects of rough solid walls.
 While several terms in a
 kinetic energy balance equation are not known experimentally, still plausible
 estimates can be performed by means of other studies and recent results for
 {an effective (isotropic)} eddy kinematic viscosity. Our results support the inference of \citet{ekm1}
 for the same quantity that it  is around  $10^{-2}$ m$^2$/s along the shorelines,
 which is several orders of magnitude smaller than the usual values used in
 computer simulations.

\section{Data and Methodology}

\subsection{Oceanic flow fields}

Recently, \citet{Risien15} compiled a set of fields of sea level anomalies by 
combining gridded daily altimeter fields with coastal tide 
gauge data (for more details see Appendix A).
\begin{figure}
 \centering
\noindent\includegraphics[width=0.47\textwidth]{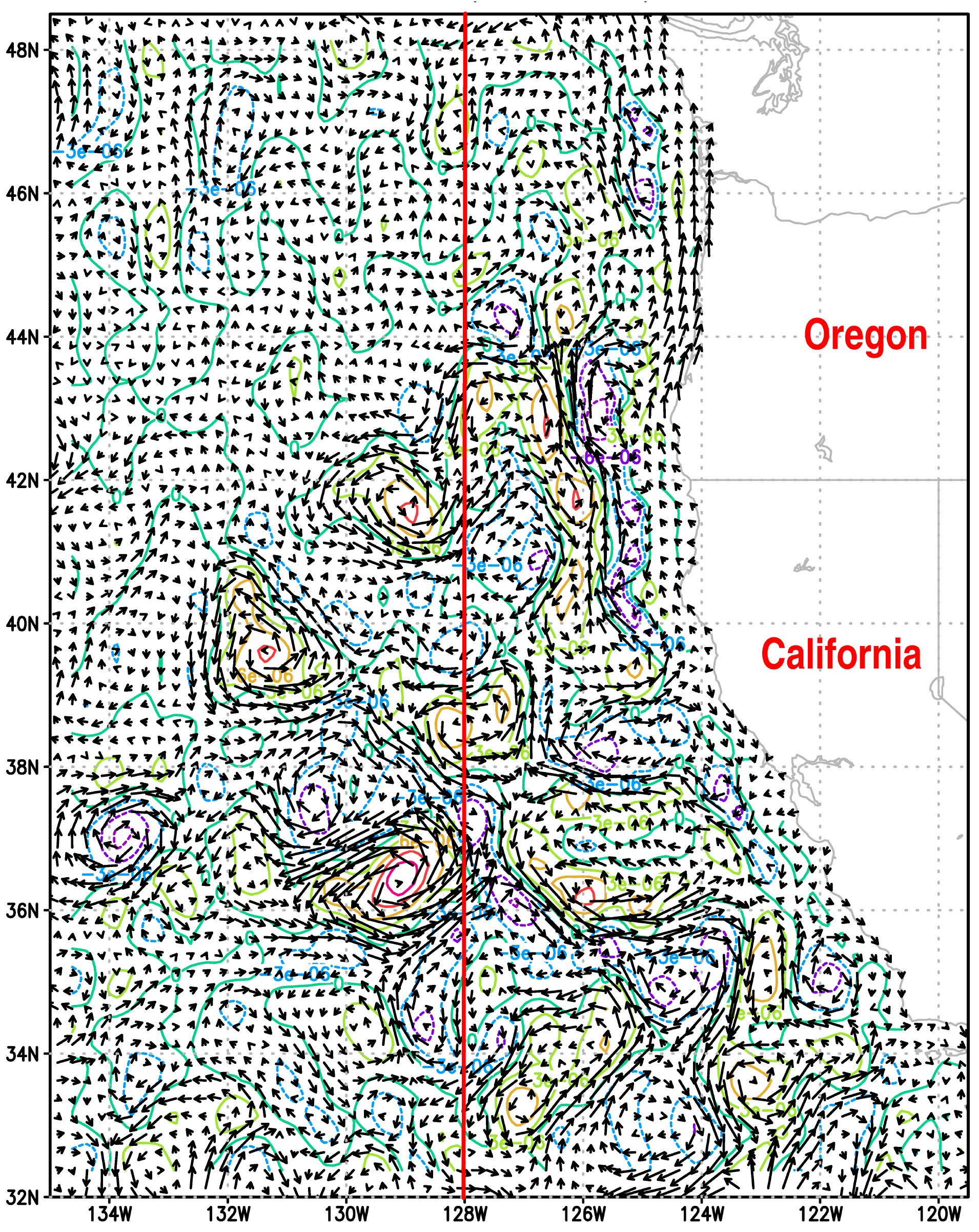}
\caption{Snapshot of the geostrophic flow field (07/01/2014) from the 
data set compiled by \citet{Risien15} along the U.S. West Coast. The longest 
arrows denote velocity vectors of 0.5 m/s, thin contour lines indicate cyclonic 
(green) and anticyclonic (blue) vorticity. The vertical red line along 
128$^\circ$W longitude separates the shore (S) and open water (O) regions for 
the subsequent analysis.}
\label{fig:1}
\end{figure}
Figure \ref{fig:1} illustrates a snapshot of the surface flow along the U.S. 
West Coast
{  obtained from sea level anomalies by assuming geostrophic balance of
hydrostatic pressure gradient and the Coriolis force}.
It is apparent that the region near the shoreline consists of 
more geostrophic vortices than away from the coast. In order to 
separate the coastal domain from the open water, we divided the area
into two parts, see the vertical red line in Fig.~\ref{fig:1}.
 The near-shoreline region eastward to
128$^\circ$W longitude consists of 1391 grid cells (3.353.407  km$^2$), while the open
ocean part westward to 128$^\circ$W is somewhat larger (1914 grid cells, 4.499.932  km$^2$).

Appropriate
macroscopic quantities characterizing the time evolution of the two dimensional flow field
are the integrated enstrophy per unit area,  $\textrm{Ens}_A(t)$, and the integrated surface kinetic energy 
per unit area, $\textrm{Ek}_A(t)$:
    \begin{equation}
	  \textrm{Ens}_A(t)=\frac{\varrho_{ref}}{2A}\iint\omega^2 dA \enspace,
	  \label{Eq:En}
    \end{equation} 

    \begin{equation}
	\textrm{Ek}_A(t)=\frac{\varrho_{ref}}{2A}\iint\mathbf{u}^2 dA \enspace,
	\label{Eq:Ek}
    \end{equation}
where $\varrho_{ref}$ is a reference density of sea water (1025 kg/m$^3$),
$\omega=\nabla\times\mathbf{u}$ is the vorticity  of the
horizontal flow field  $\mathbf{u}=(u,v)$, and $A$ 
denotes the area of surface domain of integration. The incorporation of the reference density $\varrho_{ref}$ seems
to be unnecessary, however we will use it for subsequent analysis of the total energy balance.
In this way, the dimensions of $\textrm{Ens}_A(t)$ and $\textrm{Ek}_A(t)$ are N/m$^4$ and N/m$^2$, respectively.
Normalization by the area permits to compare these quantities over the different domain sizes.

As for the numerical integration of Eq.~(\ref{Eq:En}) and Eq.~(\ref{Eq:Ek}), we consider
a given velocity component to be representative for the whole calendar day and for the
grid cell of  0.25$^\circ\times$0.25$^\circ$, where the coordinates are centered in
the cell. Grid point distances for the centered numerical derivation in $\nabla\times\mathbf{u}$ 
and grid cell areas computed by the standard approximation of a spherical Earth.
The geographic distribution of long time mean values are shown in Figs.~\ref{fig:S1} and \ref{fig:S2} in Appendix A.

\subsection{Wind stress fields}

In order to study one of the main driving forces of oceanic surface flow, we
evaluated the turbulent surface stress from the ERA-Interim reanalysis 
of the European Centre for Medium-Range Weather Forecasts (http://www.ecmwf.int/).
\citet{Brunke11} compared several products of ocean surface fluxes from
various sources and concluded that the ERA-Interim dataset is among the best
performers considering differences between observations and data bank records, also for wind stress.
Zonal and meridional  wind stress components $\vec{\tau}=(\tau_u,\tau_v)$
are  extracted from short-range forecasts and given as accumulated values with a time step of 3 hours \citep{Berrisford11}.
Since forecasts are twice daily, from 00 and 12 UTC, 
a daily mean value (properly covering diurnal cycles) is obtained  by combining two twelve hour segments of the forecasts,
over the same grid points as the surface velocity field with 0.25$^\circ\times$0.25$^\circ$ resolution.

Besides tidal forcing and the {  conversion of potential to kinetic energy via baroclinic instability}, the wind stress interacting with
surface waves, geostrophic and ageostrophic flow provides the kinetic energy for the upper ocean \citep{FeWu09, oc10}.
{  Global estimates for the world's oceans are around 3.5, 1.1, 60, 1.1 and 2.4 TW (1 teraWatt = 10$^{12}$ Watt), respectively \citep{StaWu99,FeWu04,FeWu09, oc10,Brunke11,Wu15},
see also Appendix C. (Note that far the largest fraction generates the surface waves field, however this is  almost entirely dissipated
in the surface layer turbulence and mixing.)}
 At the scales of geostrophic motions, the ocean surface is approximately horizontal, and the working rate
of wind stress per unit area (or geostrophic wind flux input)  is approximately
\begin{equation}
	\textrm{F}_A =\frac{1}{A}\iint \vec{\tau} \cdot \mathbf{u} dA \enspace,
	\label{Eq:W}
 \end{equation}
where $\mathbf{u}$  is the surface geostrophic flow in the ocean far enough from the equator,
and $A$ denotes the area of the surface domain of integration, again. This flux is given in units of  W/m$^2$.

\subsection{Numerical simulations}

The so called Lattice Boltzmann method  (LBM)  has proven to be specifically well suited for numerical studies
of rigid wall bounded 2D turbulent flows \citep{LBbook0,LBbook1}. \citet{Toth15} have recently extended 
previous LBM studies \citep{Hazi10,Toth10,Hazi11,Toth14} on the interactions between freely decaying vortices and
solid rough walls, therefore we can skip most technical details and restrict ourselves to the specific
parameters of the presented simulations.
We think that LBM simulations of freely decaying 2D turbulent flows provide a unique tool to 
clearly understand the physical link between vorticity generation and kinetic energy dissipation.
{  Several observations on the vertical velocity profile support
that the flow is nearly independent of depth within the oceanic mixed layer for time intervals
near the inertial period \citep[p. 142]{oc09}. Therefore we adopt a simple slab-flow model of uniform
velocities in the top mixed layer, which can be easily compared with 2D simulations.
}

\begin{figure*}
 \centering
\noindent\includegraphics[width=0.99\textwidth]{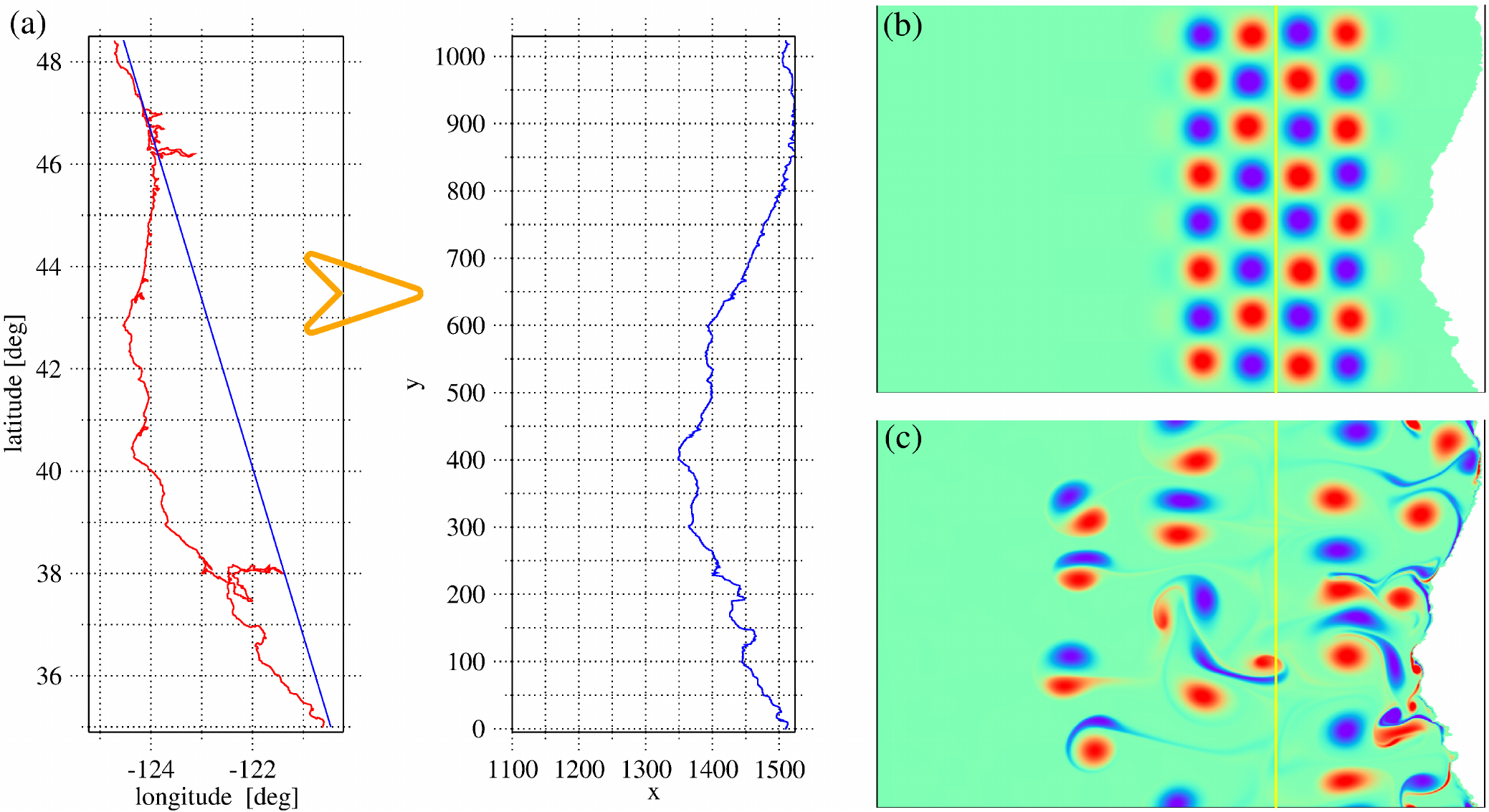}
\caption{(a) Transformation of the map of the shore for a rough line as
a solid boundary (periodic in the vertical direction) for LBM simulations. 
(b) A given initial configuration, where equal shielded Gaussian vortices of opposite signs are slightly displaced
from a regular lattice position. Vertical yellow line separates the ``shore-region'' from the ``open-water'' domain.
(c) Vorticity field after 6$\times$10$^4$ simulation steps. Red/blue color codes positive/negative vorticity.}
\label{fig:2}
\end{figure*}

Figure \ref{fig:2}a demonstrates the simple geometric procedure, where the shoreline coordinates 
are transformed to be a solid boundary of a rectangular simulation domain (1524$\times$1024 sites),
which is periodic in the vertical direction. The wall on the left side is a smooth solid line, where no-slip
boundary condition is imposed, similarly to the rough wall on the right. The flow field is open and periodic 
in the perpendicular direction. The initial configuration is a perturbed lattice of 8$\times$4 
shielded Gaussian vortices \citep{vort1,vort3,vort2}, where the center coordinates are randomly
displaced with small distances (Fig.~\ref{fig:2}b). 

The main results of the recent and very similar previous simulations with almost the same configurations \citep{Toth15} 
are the following.
\textit{(i)} Since the viscosity is finite  and there is no external driving, the total kinetic energy $\textrm{Ek}_A(t)$  given by 
 Eq.~(\ref{Eq:Ek}) is monotonously decreasing in time.
\textit{(ii)} The interaction between the vortices and the rough no-slip wall produces excess
vorticity in the flow field, which is reflected in sometimes increasing total enstrophy $\textrm{Ens}_A(t)$ 
given by Eq.~(\ref{Eq:En}).
Local enstrophy production enhances kinetic energy dissipation in the same time intervals.
\textit{(iii)} In the presence of solid walls, the following simple balance equation \citep{bal80,bal12} does not hold:
  \begin{equation}
	\frac{d\textrm{Ek}_A}{dt}=-2\nu \textrm{Ens}_A\enspace,
	\label{Eq:dEk1}
  \end{equation}
where $\nu$ is the prescribed kinematic viscosity. However, simulations with different viscosities prove
that the above relationship is correct asymptotically, when all the vortices are advected
far enough into the open flow field,  {  thus
the strong shear flows induced by the no-slip boundary condition at the wall ceased  \citep{Toth15}.
}

{  An important lesson} of the recent simulations is that the balance equation Eq.~(\ref{Eq:dEk1}) cannot
be evaluated neither locally nor asymptotically in time, when the area of integration does not cover
the entire flow field of all active vortices. In a restricted domain with open boundaires, the advection of  kinetic energy and enstrophy
always produces contributions of both signs, therefore the balance Eq.~(\ref{Eq:dEk1}) practically never holds,
even without any driving force or other dissipation mechanisms than viscosity.

\section{Results and Discussion}

\begin{figure*}
 \centering
\noindent\includegraphics[width=0.98\textwidth]{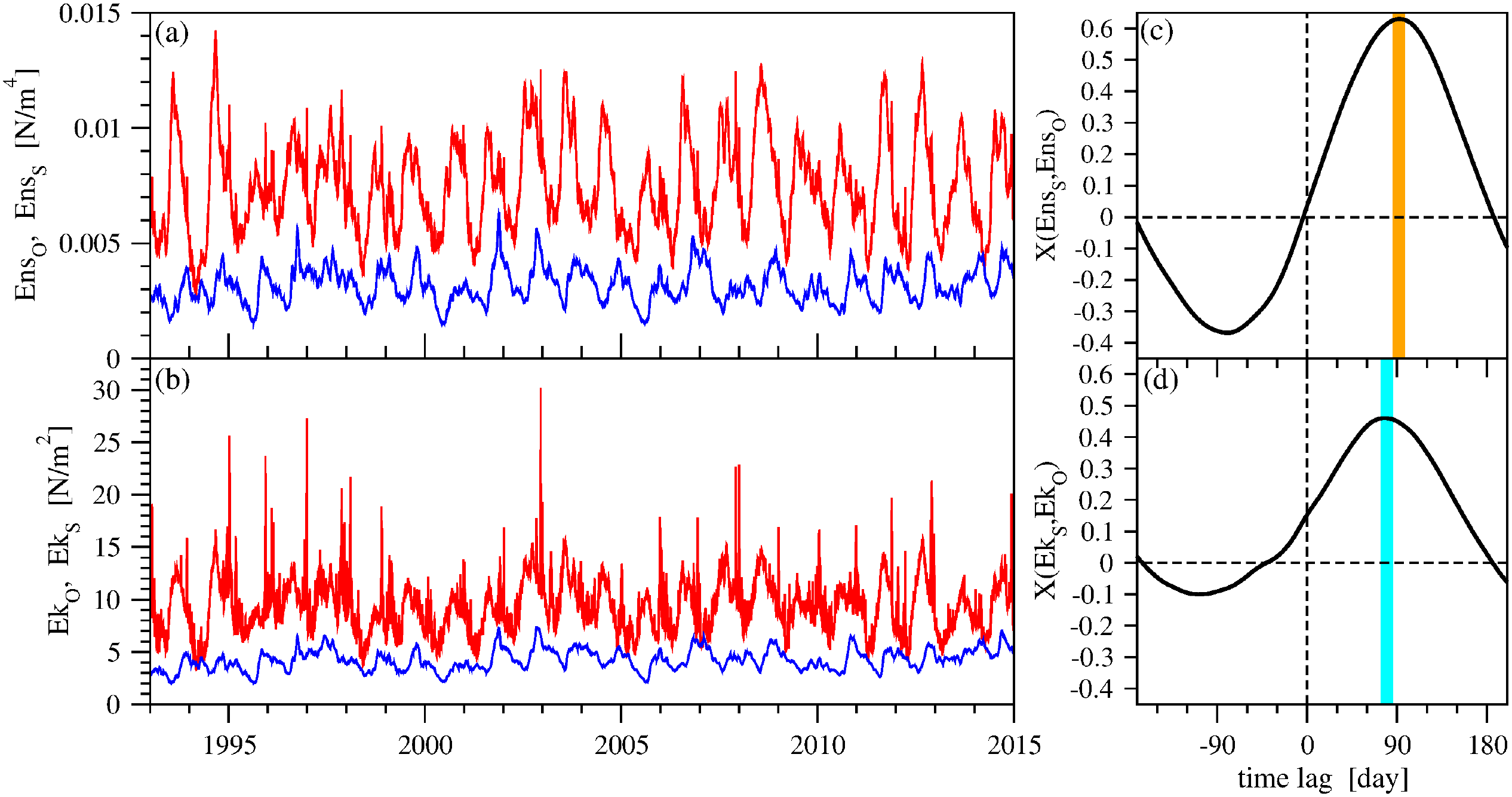}
\caption{
(a) Integrated 2D enstrophy per unit area [see Eq.~(\ref{Eq:En})] for the open water
($\textrm{Ens}_O(t)$, blue) and for the shore ($\textrm{Ens}_S(t)$, red) as a function of time.
(b) Integrated 2D kinetic energy per unit area [see Eq.~(\ref{Eq:Ek})] for the open water
($\textrm{Ek}_O(t)$, blue) and for the shore ($\textrm{Ek}_S(t)$, red) as a function of time.
(c) Cross correlation [see Eq.~(\ref{Eq:X})] for $\textrm{Ens}_S(t)$ and 
$\textrm{Ens}_O(t)$, the time lag of the maximum correlation (88-98 days) indicated
by an orange stripe.
(d) Cross correlation [see Eq.~(\ref{Eq:X})] for $\textrm{Ek}_S(t)$ and 
$\textrm{Ek}_O(t)$, the time lag of the maximum correlation (75-85 days) indicated
by a cyan stripe.}
\label{fig:3}
\end{figure*}

In order to analyze the role of coastal processes, the integrated quantities given by
Eqs.~(\ref{Eq:En}) and (\ref{Eq:Ek}) are determined for the two regions separated
by the vertical red line in Fig.~\ref{fig:1}, the results are  plotted in Figure \ref{fig:3}.
It is evident for the 2D enstropy per unit area and 2D kinetic energy per unit area that both are
much larger in the coastal region than in the open water (see also Figs.~S1 and S2). Both quantities exhibit a
marked annual oscillation  indicating that wind driven
processes play the main role in the production of vorticity and kinetic energy. 

 A simple visual comparison of the curves suggests that increasing and decreasing phases
of enstrophy and kinetic energy over the open water domain develop after a time
delay with respect to the coastal area. The standard way of determining such delay
between two stationary  time series $x(t)$ and $y(t)$ is given by the cross correlation
function:
\begin{equation}
\textrm{X}_{x,y}(\Delta)= \frac{\langle [x(t)-\bar{x}][y(t+\Delta)-\bar{y}]\rangle_t}{\sigma_x\sigma_y} 
\enspace ,
\label{Eq:X}
\end{equation}
where $\Delta$ is the time lag (can be negative), $\langle\cdot\rangle_t$ denotes
averaging over time,  $\bar{x}$  and $\bar{y}$ are temporal mean values, 
furthermore $\sigma_x$ and $\sigma_y$ are the standard deviations
of the two time series. The results are plotted in Fig.~\ref{fig:3}c for enstrophy and
in Fig.~\ref{fig:3}d for kinetic energy. The maxima seem to be significant at around a
delay of 3 months suggesting that  {  advection  from the near-shore
region is an important mechanism besides local generation of vorticity and kinetic energy
over the open water domain \citep{drift1,drift2,Chelton11}.} The characteristic delay of 90 days is consistent with the 
results of \citet{Chelton11} on
the mean westward propagation speed of mesoscale eddies. They obtained around 0.03 m/s in the
latitude band of 30$^\circ$N-40$^\circ$N (see Fig.~22 in \citet{Chelton11}) which is equivalent
with 233 km propagation distance during 90 days. This distance is actually the active width of the shore
region (255 km between 125$^\circ$W and 128$^\circ$W longitude along 40$^\circ$N).

The time evolution of the kinetic energy can be derived from theoretical considerations
of the total energy budget of the oceans \citep{bal80,CS04,FeWu09,bal12,Pond13}. 
{  We adopt the simplest model for turbulence closure, where the Reynolds stress is proportional 
to the deformation rate tensor, and one can introduce an isotropic turbulent viscosity. The Reynolds-averaged 
momentum equation multiplied by the mean velocity and density provides the budget equation
 for kinetic energy tendency. When we exploit the 2D feature of the flow, the friction term (Laplacian) 
can be expressed by means of the enstrophy, see also Appendix C.
}
For an energy budget per unit area, we
need volume integrals of the kinetic energy and enstrophy:
 \begin{equation}
	\frac{d}{dt} \int_{-Z}^0 \textrm{Ek}_A dz= \textrm{F}_A -2\nu \int_{-Z}^0 \textrm{Ens}_A dz + S_A\enspace,
	\label{Eq:dEk2}
  \end{equation}
where $\textrm{Ek}_A$, $\textrm{Ens}_A$ and $\textrm{F}_A$ are given by Eqs.~(\ref{Eq:Ek}), 
(\ref{Eq:En}) and (\ref{Eq:W}), $\nu$ is  {  an effective (turbulent) kinematic viscosity} of the sea water (in units of m$^2$/s), and
$dz$ indicates integration in the vertical direction.  {  It is important that $\nu$ is a property of the flow field, not of the fluid.}
The last term in Eq.~(\ref{Eq:dEk2}) $S_A$ contains 
all other sources such as tidal forcing, wind stress interacting with ageostrophic flow, advection,
{  and conversion of potential energy through baroclinic instability},
which we cannot extract and separate from the available data. Note that the depth of vertical integration $Z$
 in  Eq.~(\ref{Eq:dEk2})  is {  not known a priori}, using numerical values from  Eqs.~(\ref{Eq:En}) and 
(\ref{Eq:Ek}) assumes implicitly a depth of 1 m, which is certainly a serious underestimation
of the reality, see below. 

\begin{figure*}
 \centering
\noindent\includegraphics[width=0.98\textwidth]{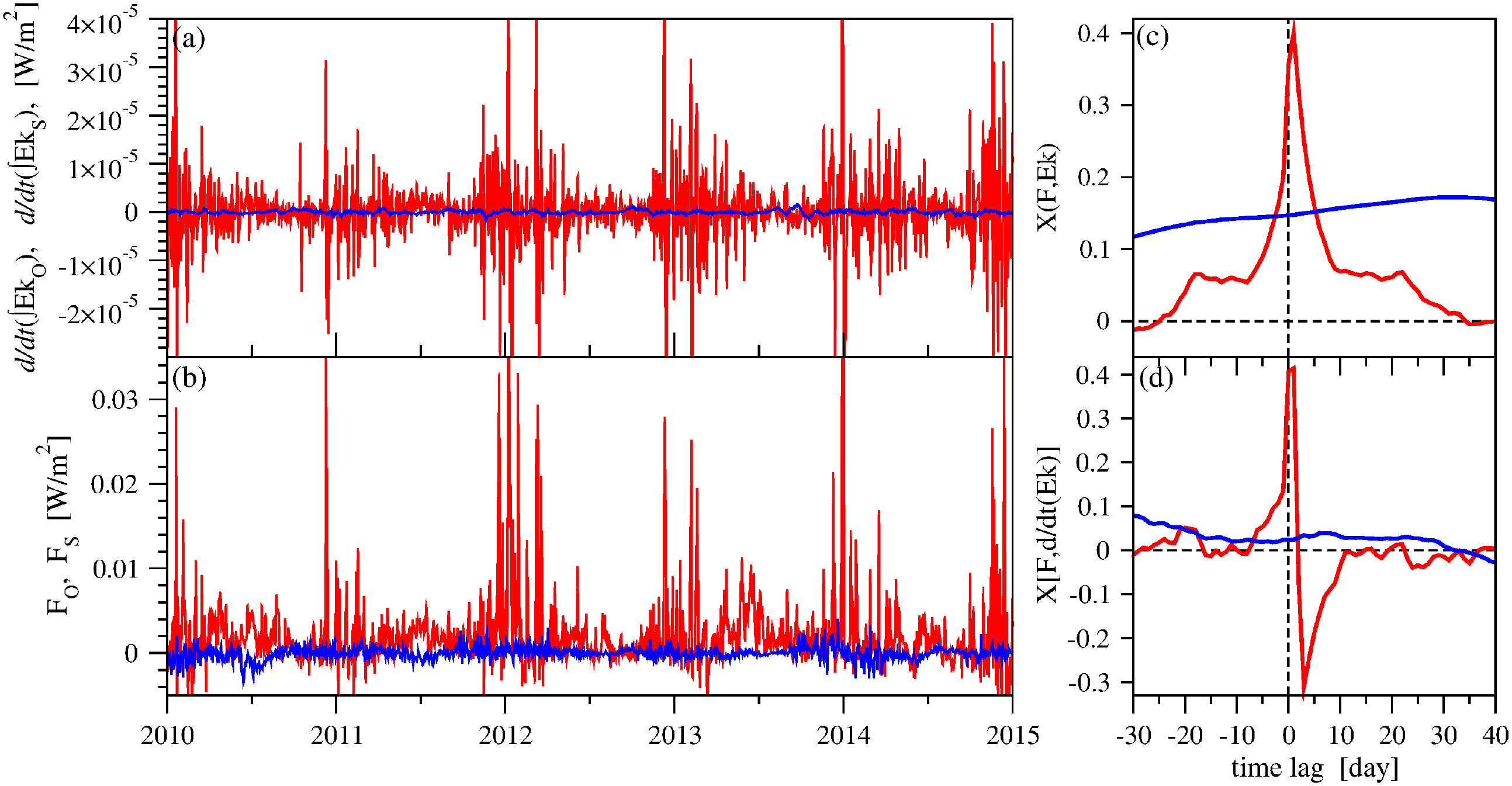}
\caption{
(a) {  Kinetic energy tendency} [see Eq.~(\ref{Eq:dEk2})] for the open water
(blue) and for the shore region (red) as a function of time (the last five years are shown).
(b) Wind stress work [see Eq.~(\ref{Eq:W})] for the open water
(blue) and for the shore region (red) as a function of time.
(c) Cross correlation [see Eq.~(\ref{Eq:X})] for the wind stress work $\textrm{F}(t)$ and 
the kinetic energy $\textrm{Ek}(t)$, blue line is for the open water, red is for the shore domain.
(d) Cross correlation [see Eq.~(\ref{Eq:X})] for the wind stress work $\textrm{F}(t)$  and 
the surface {  kinetic energy tendency} [lhs of Eq.~(\ref{Eq:dEk2})],
 blue line is for the open water, red is for the shore domain.}
\label{fig:4}
\end{figure*}

Figures \ref{fig:4}a and \ref{fig:4}b illustrate the relationship between the wind stress energy flux and the time derivative
of kinetic energy per unit area ({  kinetic energy tendency}) both for the open water section (blue lines) and for the shore region (red lines).
{  The comparison of the vertical scales in Figs.~\ref{fig:4}a and \ref{fig:4}b indicates  that the net kinetic 
energy input is almost negligible in relation to the wind stress work}, especially over the open water region. 
Indeed, the mean magnitude ratio of the two quantities is around
$1.78\times 10^{-3}$ over the shore- and $4.6\times 10^{-4}$ over the open domains.
Note that we are working with daily representative values, thus a mean daily change of kinetic energy is
$86400 \times 4.15\times 10^{-6} \approx 0.36$ J/m$^2$,
while the mean magnitude of total daily wind work is 
$86400 \times 2.33\times 10^{-3} \approx 202$ J/m$^2$ over the shore region, the same values
are 0.022 J/m$^2$ and  46.8 J/m$^2$ over the open ocean  ($Z=1$ m).

{   The vertical integration
affecting both the kinetic energy tendency and enstrophy is a simple multiplication in the slab-flow model
approximation, still the proper choice of the depth $Z$ is not easy. Values from 2 m from the surface  down to hundreds
of meters are reported from various observations and modeling, we discuss this problem in Appendix D.
At the end we consider two integration depths for the final estimates of the terms  in Eq.~(\ref{Eq:dEk2}),
$Z\approx 20$ and 50 m (we think that larger values, e.g. 200 m, would be a serious overestimate of a possible slab-flow feature)}.

Empirical cross correlation relationships between the wind stress work and kinetic energy, and
wind stress work and {  kinetic energy tendency} are plotted in Figs.~\ref{fig:4}c and \ref{fig:4}d. Red/blue curves
denote the shore/open water region. It is quite obvious that kinetic energy has a definite real time cross correlation
with a maximal delay of 1 day in the shore region, i.e. the response arrives almost synchronously with the changes of wind forcing.
Such cross correlation does not appear in the open water region. The most probable reason is that the characteristic
horizontal size of geostrophic eddies is in the range of 50-200 km, however the smooth wind field over the
open oceans has a much higher correlation length  \citep{wind16}, which is typically 300-600 km even over land \citep{wind15}. 
Wind blowing over one side of a geostrophic eddy can accelerate the flow, however
on the opposite side of a closed eddy it has a decelerating effect.
Long time local mean values for the wind stress work are also very small over open water (see Fig.~S3).
{  As for the curves in Fig.~\ref{fig:4}d (cross correlation between wind stress work and kinetic energy tendency), the
negative value at a time lag of 3-4 days is a direct consequence of the oscillatory nature of the
kinetic energy tendency, only for the shore region. Since the characteristic autocorrelation
time of the wind field is 3-5 days, a strong 
stroke of kinetic energy input is regularly followed by a drop in a couple of days due also to an enhanced dissipation
in the coastal region.}

The eddy kinematic viscosity $\nu$ in Eqs.~(\ref{Eq:dEk1}) and (\ref{Eq:dEk2}) in numerical simulations prescribed
in a rather heuristic way. A recent review by \citet{nu16} lists 14 works where the minimal range
of  $\nu$ is 50-100, while the maximal range is as large as 6000-10000  m$^2$/s.
Here we attempt to estimate an effective $\nu$  based on our empirical data.

Firstly, Eq.~(\ref{Eq:dEk1}) holds asymptotically in the 2D LBM simulations, see Fig.~\ref{fig:S4}a in Appendix B.
However, when the area of integration is restricted to the ``open water'' region
(left to the yellow line in Figs.~\ref{fig:2}b and \ref{fig:2}c), the {  kinetic energy tendency} per unit area 
becomes often positive indicating inward advection of vortices, and Eq.~(\ref{Eq:dEk1}) does not
hold (Fig.~\ref{fig:S4}b).  {  The method of fitting an envelop (see Fig.~\ref{fig:S4}a) does not work for the empirical 
flow field either (Fig.~\ref{fig:S4}c), further complicated with the fact that
other source terms than advection, such as  baroclinic instability possibly
give essential contributions to the geostrophic flow field.
}

Secondly, based on the observation that the {  kinetic energy tendency is orders of magnitude smaller than the wind stress work,
even for $Z=50$ m, we can neglect it in Eq.~(\ref{Eq:dEk2}), and get the energy budget equation  representing a quasistationary
dynamics}:
 \begin{equation}
	 \textrm{F}_A + S_A = 2\nu \int_{-Z}^0 \textrm{Ens}_A dz \enspace,
	\label{Eq:dEk3}
  \end{equation}
[notations are identical with  Eq.~(\ref{Eq:dEk2})]. 
Figure S5 helps in the final estimate  by illustrating together two terms in Eq.~(\ref{Eq:dEk3}), namely the
wind stress work per unit area $\textrm{F}_A$ (black lines) and the depth integrated enstrophy per unit area
(here the implicit integration depth is  $Z=1$ m) for the shore (red) and open water (blue) domains.
Note again the disparity between the shore and open water regions, especially considering the wind stress work:
the long time mean value of this forcing over the open water is practically zero (black curve in Fig.~S5, and also Fig.~S3).
For this reason, in the final calculation we consider the long time mean of the absolute value, $\vert \bar{\textrm{F}}_{O} \vert = 2.8\times 10^{-4} $ W/m$^2$.
The main unknown in  Eq.~(\ref{Eq:dEk3}) is the compound source $S_A$ incorporating 
tidal forcing,  {   conversion of potential to kinetic energy via baroclinic instability}, ageostrophic flow, advection, etc.,
see Subsection 2.2 and Appendix C.
Since our goal is an order of magnitude estimate of $\nu$, we consider two limiting cases: (i)  $ S_A\approx\textrm{F}_A$ (wind stress work is dominating),
and (ii) $ S_A\approx 10\cdot \textrm {F}_A$ (wind stress work is only 10 \% of the total forcing). 
 {  The final results for $\nu$ by using $Z = 20$ and 50 m integration depths are shown in Table 1 in Appendix D.
The same evaluation can be performed separately for each grid cell, the result is shown in Fig.~S6.
Apart from noise, the tendency is similarly clear, an effective eddy coefficient of viscosity $\nu\approx 10^{-2}$ m/s$^2$ 
decreases in the offshore regions by a factor 2.}

\section{Conclusions}

We can summarize our main findings as follows. (i) The coastal region is an {  essential} source of kinetic energy and enstrophy
(geostrophic vortices), and they are advected westward to the open water. (ii) Results from the 2D Lattice Boltzmann
simulations (freely decaying vortices interacting with a solid rough wall) demonstrate that excess vorticity is generated
along the shoreline, while kinetic energy decreases monotonously without external forcing. The 2D energy tendency - dissipation
balance [Eq.~(\ref{Eq:dEk1})] does not hold in a finite open region because of advection effects. (iii) Wind forcing 
(wind stress work on geostrophic vortices) is negligible in the open water region. (iv) {  Kinetic energy tendency} is
extremely small everywhere compared with wind stress work  {    (see Figure \ref{fig:S7} in Appendix D)}, thus the 
geostrophic flow can be considered
as quasistationary on the timescale of a few days. (v) An evaluation of the kinetic energy budget equation
 [Eq.~(\ref{Eq:dEk3})] results in a robust estimate of an effective eddy kinematic viscosity in the order of magnitude $10^{-2}$ m$^2$/s,
 where the numerical values are systematically lower in the open water region.

\begin{acknowledgments}
{  Geostrophic sea surface velocities compiled by \citet{Risien15} can be accessed at https://ir.library.oregonstate.edu/xmlui/handle/1957/57170.
ERA Interim daily data can be downloaded after a simple registration at http://apps.ecmwf.int/datasets/data/interim-full-daily/levtype=sfc/.}
This work is supported by the Hungarian National Research, Development and Innovation Office under grant numbers 
FK-125024 and K-125171.
\end{acknowledgments}

\appendix

\section{ Methodolgy of compiling the geostrophic flow field}

\begin{figure}
\centering
\noindent\includegraphics[width=0.47\textwidth]{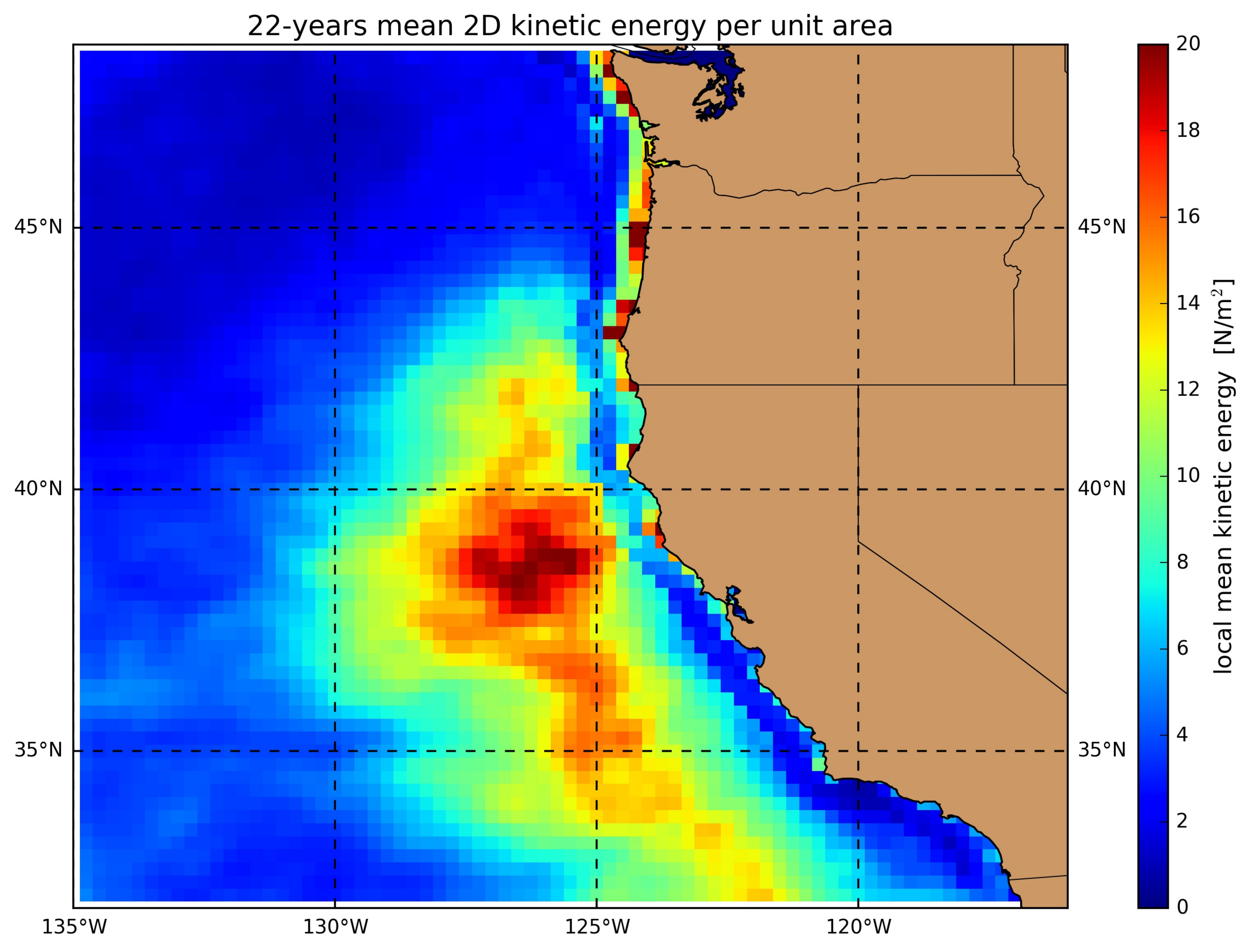}
\caption{Geographic distribution of the local mean kinetic energy averaged over the 22 years of daily records.
}
\label{fig:S1}
\end{figure}

\begin{figure}
\centering
\noindent\includegraphics[width=0.47\textwidth]{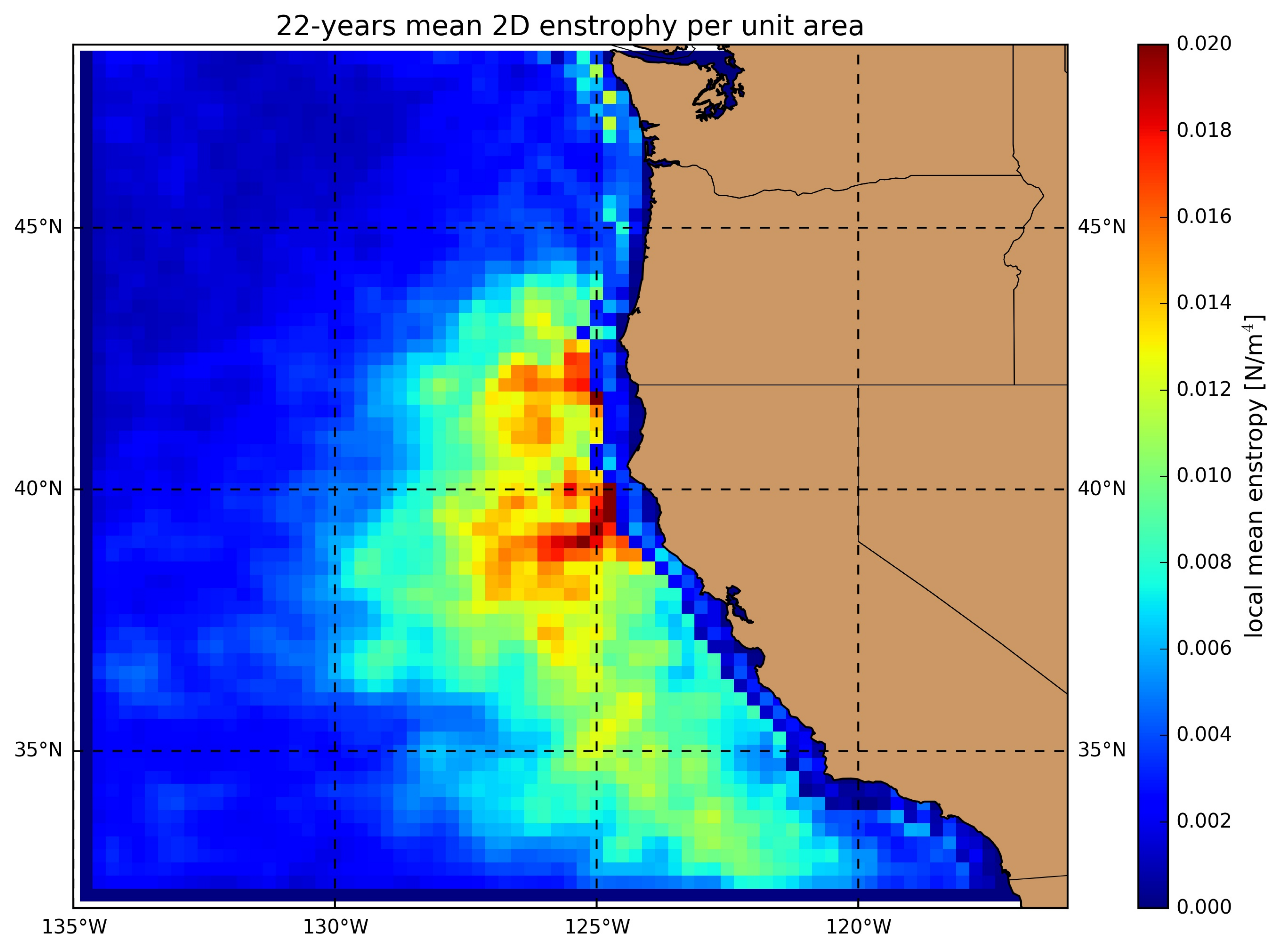}
\caption{Geographic distribution of the mean enstrophy (local value of squared vorticity) averaged over the 22 years of daily records.
}
\label{fig:S2}
\end{figure}

The geographic area covers  32.0$^\circ$N -- 48.5$^\circ$N (latitude) and 
135.0$^\circ$W -- 111.25$^\circ$W (longitude) with a spatial resolution of 
0.25$^\circ\times$0.25$^\circ$.
Within approximately 50-80 km of the coast, the altimeter data are discarded and 
replaced by a linear interpolation between the tide gauge and 
remaining offshore altimeter data \citep{Saraceno08}. A 20-year mean is 
subtracted from each time series (tide gauge or altimeter) before combining the 
data sets to 
form the merged sea level anomaly data set. Geostrophic velocity anomaly fields 
are formed from the surface heights. Daily mean fields are produced 
for the period 1 January 1993 - 31 December 2014. The primary validation 
compares geostrophic velocities calculated from the height fields and 
velocities measured at four mooring sites covering the north-south range of the
data set \citep{Risien15}. Figures \ref{fig:S1} and \ref{fig:S2} illustrate the geographic distribution of long time mean values
for the kinetic energy and enstrophy determined separately in each grid cell (see main text).

\section{Asymptotic method to estimate an eddy kinematic viscosity}

\begin{figure*}
 \centering
\noindent\includegraphics[width=0.98\textwidth]{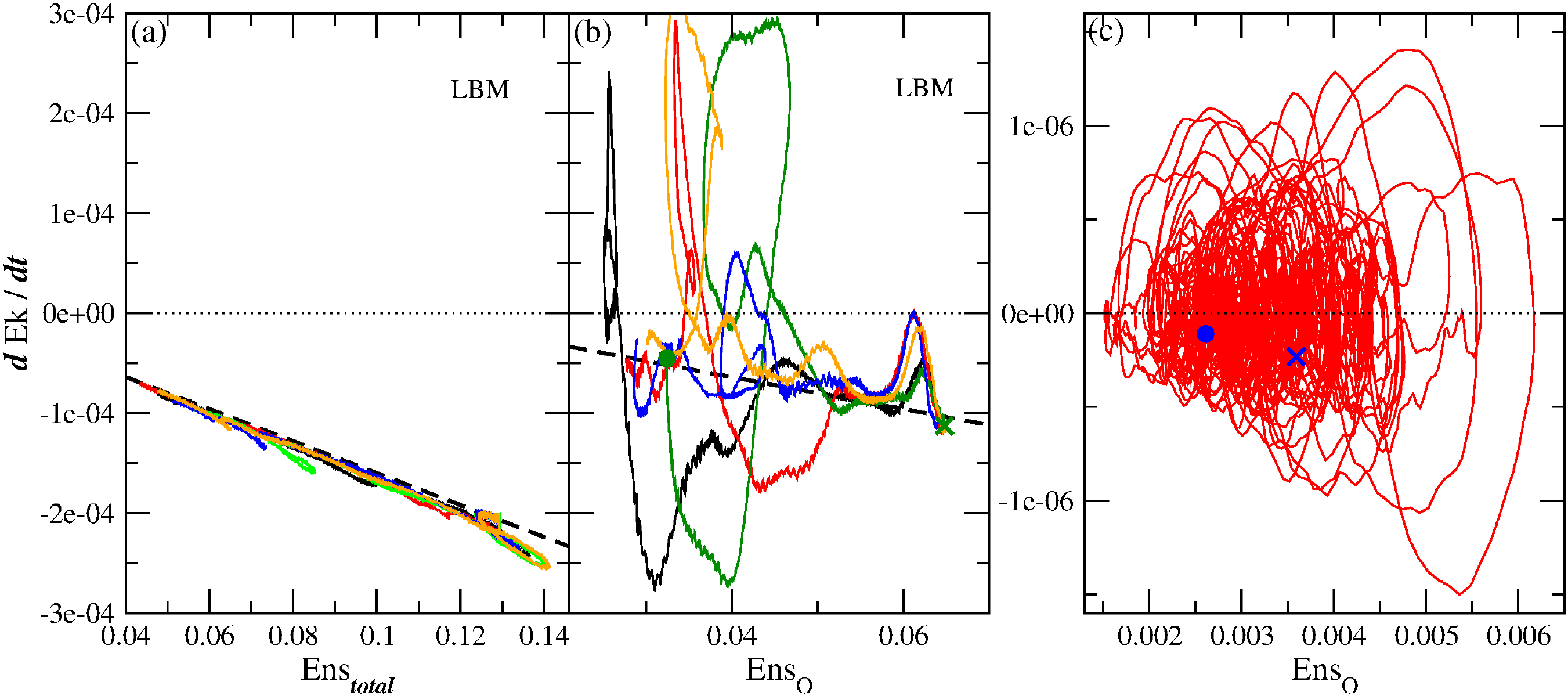}
\caption{Energy balance relationship  [see Eq.~(4) in the main text] for LBM simulations and the empirical data.
(a) Time derivative of the kinetic energy as a function of total enstrophy Ens$_{total}$
[integration in Eq.~(1) is the total area containing all the vortices], LBM simulations.
Different colors denote different realizations from preturbed initial configurations, the dashed line
is the expected slope of $-2\nu$ (not fitted).
(b) The same as (a), however the analysis is restricted in the ``open water'' area (see Fig.~2 in the min text,
left from the yellow line). Heavy dots an crosses indicate the first and last values of the
corresponding records.
(c) The same analysis on empirical data in the ``open water''  region  (see Fig.~1 in the main text,
left from the red line). Heavy dot/cross indicates the first/last point in the time series.
Intrinsic noise is damped by running mean smoothing in each panel.
}
\label{fig:S4}
\end{figure*}

Firstly, Eq.~(4) holds asymptotically in the 2D LBM simulations, illustrated properly in Fig.~\ref{fig:S4}a.
Here the kinetic energy tendency is plotted as a function of area integrated enstrophy. It is essential that the
integrated area extends to the whole simulation domain including all the vortices. The characteristic loops
indicate enhanced dissipation as a consequence of vortex-wall interaction (excess shear as a consequence of
no-slip boundary condition), however the envelope of the
curves produced from different random initial conditions perfectly coincides with the expected value of $-2\nu$.
The bad news is in  Fig.~\ref{fig:S4}b: when the area of integration is restricted to the ``open water'' region
(left to the yellow line in Figs.~2b and 2c, see the main text), the kinetic energy flux per unit area 
becomes often positive indicating inward advection of vortices, and Eq.~(4) does not
hold. An attempt to perform the same analysis in the measured surface flow data remains
inconclusive too (Fig.~\ref{fig:S4}c illustrates the result for the open water area). The extension of the
area to the whole test region does not help, because the domain remains open where drifting in and out
of vortices is a dominating part of the dynamics.

\section{Derivation of energy balance equations in 2D  slab-flow approximation}

In our approach, let us consider the well known Navier-Stokes momentum balance equation in a shallow layer of depth $Z$,
where $\mathbf{u}=(u,v,0)$ is the horizontal uniform velocity with zonal ($u$) and meridional ($v$) components.
This arrangement is the slab-flow model of ocean surface currents in the top mixed layer of constant reference density 
$\varrho_0$:
\begin{equation}
\frac{\partial\mathbf{u}}{\partial t}+\left( \mathbf{u}\nabla\right)\mathbf{u} = -f\mathbf{n}\times\mathbf{u}-
\frac{1}{\varrho_0}\nabla p + \nabla\left(\nu_m\nabla\mathbf{u}\right) + \mathbf{f} \enspace . \label{S1}
\end{equation}
Here
$f$ is the Coriolis parameter,  $\nabla = (\partial/\partial x, \partial /\partial y, \partial /\partial z)$ 
is the nabla operator, $p$ is the pressure, $\nu_m$ 
is the molecular kinematic viscosity, and $\mathbf{f}$ contains all forcing terms (force per unit mass) .
The incompressibility constraint is expressed as $\nabla \mathbf{u} = 0$.

Reynolds decomposition applied to the Navier-Stokes equation is a statistical method to
describe small scale turbulent processes by separating  any flow variable into a mean part and
fluctuation around the mean, such as $\mathbf{u}=\mathbf{\bar u} + \mathbf{u}' $.
Mean values can be obtained by averaging over temporal intervals or/and spatial domains, in any case,
the mean of the fluctuations must be zero, by definition. The Reynolds-averaged momentum equation has the following form:
\begin{equation}
\frac{\partial\mathbf{\bar u}}{\partial t}+\left( \mathbf{\bar u}\nabla\right)\mathbf{\bar u} = -f\mathbf{n}\times\mathbf{\bar u}-
\frac{1}{\varrho_0}\nabla {\bar p} + \nabla\left(\nu_m\nabla\mathbf{\bar u}\right) + \mathbf{\bar f} -
\overline{ \left( \mathbf{ u'}\nabla\right)\mathbf{u'}}\enspace .  \label{S2}
\end{equation}
The last term in this equation is the Reynolds stress tensor, however equations for its components
would involve  new unknowns. This well known closure problem implies that if one is only interested in
the mean, the effect of the Reynolds stress components  have to be
modeled. The simplest consistent model is to introduce a turbulent kinematic viscosity $\nu_t$ as a proportionality constant between
the Reynolds stress components and the deformation rate tensor \citep[see e.g.,][]{Pond13}. Assuming isotropy both for the
molecular and turbulent coefficients of viscosity as a first approximation, 
we get back formally the pointwise Navier-Stokes
equation now for the mean velocities:
\begin{equation}
\frac{\partial\mathbf{\bar u}}{\partial t}+\left( \mathbf{\bar u}\nabla\right)\mathbf{\bar u} = -f\mathbf{n}\times\mathbf{\bar u}-
\frac{1}{\varrho_0}\nabla {\bar p} + \nu\nabla^2\mathbf{\bar u} + \mathbf{\bar f} \enspace ,  \label{S3}
\end{equation}
where the effective kinematic viscosity coefficient $\nu = (\nu_m + \nu_t)$ contains the molecular- and eddy
viscosities. Since the eddy coefficient is orders of magnitude larger than the molecular one, $\nu $ is 
usually termed as eddy kinematic viscosity.

All terms of Eq. (S3)  multiplied by $\mathbf{\bar u}$  and $\varrho_0$
will provide the tendency of mean kinetic energy per unit depth:
\begin{equation*}
\varrho_0\mathbf{\bar u}\frac{\partial\mathbf{\bar u}}{\partial t}+\varrho_0\mathbf{\bar u}\left( \mathbf{\bar u}\nabla\right)\mathbf{\bar u} = -f\varrho_0\mathbf{\bar u}\mathbf{n}\times\mathbf{\bar u}-
\mathbf{\bar u}\nabla {\bar p} + \nu\varrho_0\mathbf{\bar u}\nabla^2\mathbf{\bar u} + \varrho_0\mathbf{\bar u}\mathbf{\bar f} \enspace .  \label{S4}
\end{equation*}
Since the Coriolis force is perpendicular to the velocity, the first term on the rhs drops out.
For the same reason, in strict 2D flows gravity does not contribute to kinetic energy changes.
Next we exploit the following vector identities by using the incompressibility constraint $\nabla \mathbf{u} = 0$ in (S7):
\begin{equation}
\mathbf{\bar u}\frac{\partial\mathbf{\bar u}}{\partial t} = \frac{1}{2}\frac{\partial\mathbf{\bar u}^2}{\partial t} \enspace ,  \label{S5}
\end{equation}
\begin{equation}
\mathbf{\bar u} (\mathbf{\bar u}\nabla)\mathbf{\bar u} =\mathbf{\bar u} \frac{1}{2}\nabla \mathbf{\bar u}^2 -\mathbf{\bar u}(\mathbf{\bar u}\times \omega) \enspace, \label{S6}
\end{equation}
\begin{equation}
\mathbf{\bar u}\nabla^2\mathbf{\bar u} = - \mathbf{\bar u}(\nabla\times \omega )= - \omega(\nabla\times  \mathbf{\bar u}) = - \omega^2 \enspace ,  \label{S7}
\end{equation}
where we use the notation for the mean vorticity $\omega = \nabla \times \mathbf{\bar u}$ . Note that the first step where we
explicitly utilize the 2D feature of the flow field is in Eq.~(S6), in order to remove the second term on the rhs (the vorticity
vector is strictly perpendicular to the velocity).

When we adopt the notations of Eqs. (1) and (2) for the enstrophy and
kinetic energy per unit area (see the main text), we obtain the following budget equation from (S4):
\begin{equation}
\frac{d}{dt}\mathrm{Ek}_A=\frac{\partial}{\partial t}\mathrm{Ek}_A
+ \mathbf{\bar u}\nabla \mathrm{Ek}_A = - \mathbf{\bar u}\nabla {\bar p}
- 2\nu\mathrm{Ens}_A + \varrho_0\mathbf{\bar u}\mathbf{\bar f} \enspace .  \label{S8}
\end{equation}
Note that each term has the unit of W/m$^3$, which can be interpreted as some kind of surface flux per unit depth. 

External forces driving the ocean into motion on any scale are restricted in number 
\citep{FeWu09,oc10}. Possible forces are  winds, air-sea
exchange of sensible and latent heat, the exchange of freshwater, pressure
loading by the atmosphere, tides, geothermal heating, and  biology.
The wind field is by far the largest kinetic energy source to the ocean,
the global annual net value is estimated around 65 TW (1 teraWatt = 10$^{12}$ Watt).
Most of the energy input (cca. 60 TW)  generates the surface wave field, but this 
fraction is almost totally dissipated in the top mixed layer turbulence (wave breaking and
mixing). At the scales of geostrophic
motions, the ocean surface is approximately horizontal, and the working rate is 
given by Eq. (3) (see the main text), with a global annual net value of 0.8-1.1 TW.
The Ekman layer (ageostrophic flow) absorbs 2.4 TW input annually, a small part (0.2 TW) is
supposed to generate internal waves. The atmospheric pressure loading incorporated in the
first term on the rhs of Eq. (S8) is negligible in the kinetic energy budget (with a global value
of around 0.04 TW). This horizontal pressure gradient term includes also the conversion of potential 
energy through baroclinic instability, a global estimate is around 1.1 TW for the world’s oceans \citep{Hu04}.
 The remaining major input flux is tidal forcing which is dissipative at the bottom,
nevertheless drives the total water body having a net positive sign (input of mean kinetic energy).
The total amount of tidal forcing in the world’s oceans is estimated around 3.5 TW \citep{FeWu09,oc10} .

\begin{figure}
\centering
\noindent\includegraphics[width=0.47\textwidth]{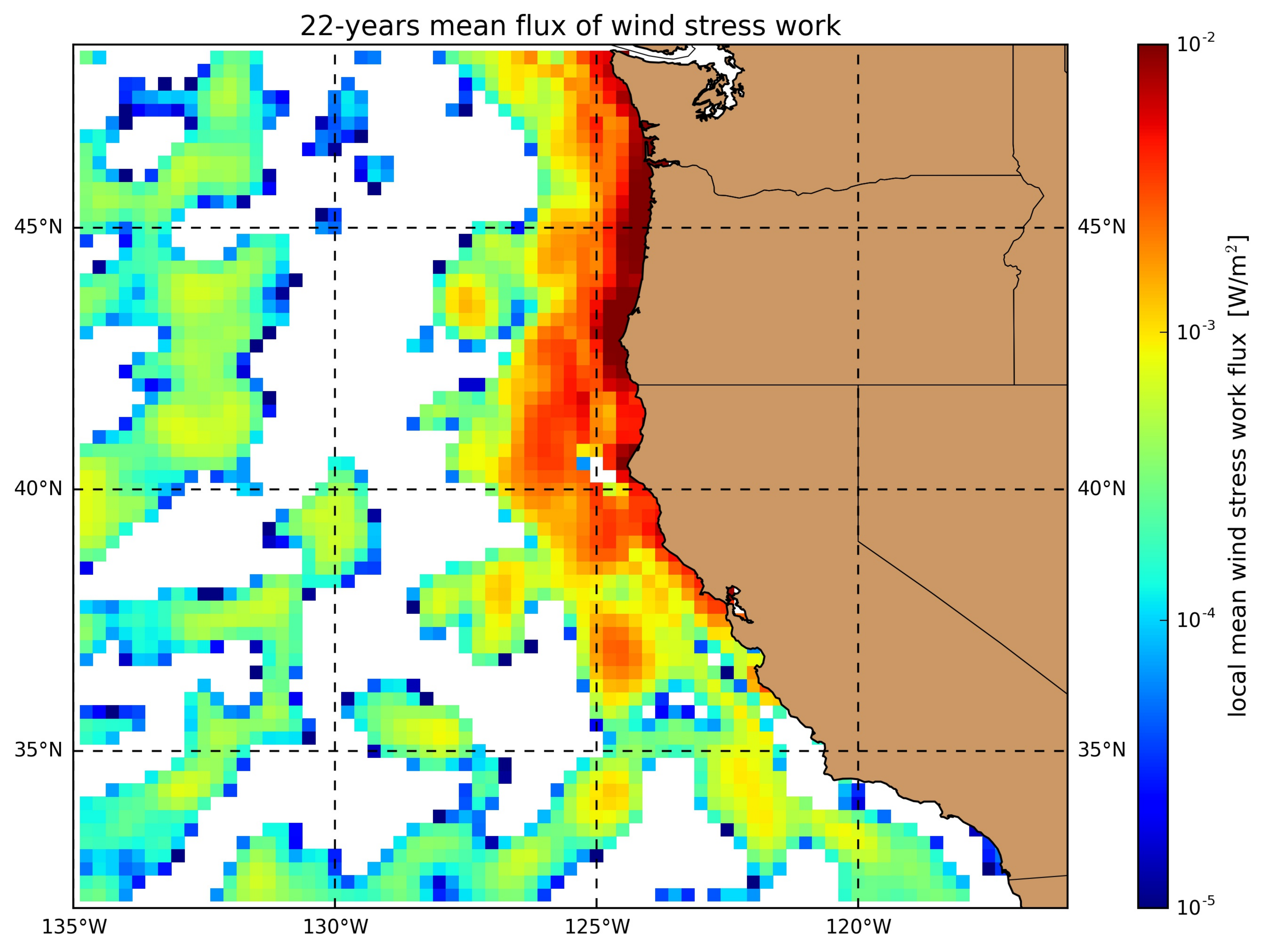}
\noindent\includegraphics[width=0.47\textwidth]{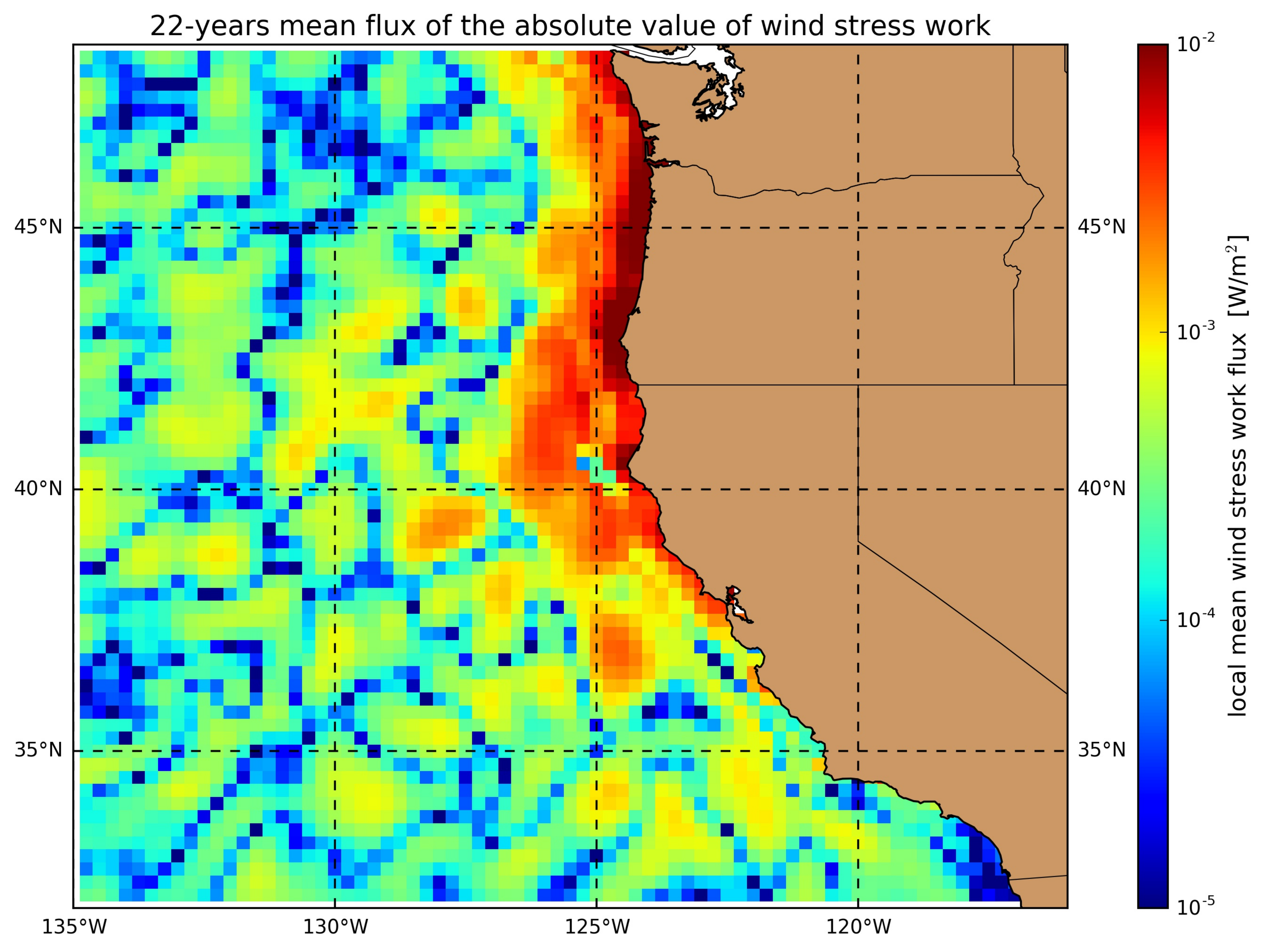}
\caption{Geographic distribution of the local mean wind stress work (top) and its absolute value (bottom) 
averaged over the 22 years of daily records.
Note that the color bar is scaled logarithmically. White areas denote very small negative values on the top panel.
}
\label{fig:S3}
\end{figure}

\section{The problem of choosing an integration depth}

 Since both the kinetic energy tendency and enstrophy are affected by the choice of integration depth $Z$,
it would be desirable to have a reliable estimate of the vertical extent of mesoscale eddies.
Unfortunately, this issue is rather controversial in the available literature. 
In the framework of our simplified model, we assume that the mixed layer moves like a slab
at least for 16-18 hours in the given region (inertial period), and current shear is concentrated at the top of the
thermocline.  Figure 7 of  \citet{ekm1} displays the vertical profile of geostrophic velocity relative to 2800 m,
when she evaluated the
surface Ekman spiral approximately 400 km off the coast of northern California (almost the middle of our test area).
 The top 40 m clearly
 reflects a uniform flow (the reported maximum mixed layer depth was also 40 m during the observing period).
A moderate value of 5 m considered as representative for the surface layer was implemented in a decomposition of 
kinetic energy in the Central Mediterranean by \citet{eddy11}.
Even lower values obtained by \citet{ekm2} who performed dye experiments and large-eddy simulations in order to
study the flow in the mixed layer, they observed mixed layer depths between 2 and 18 m, again close to the east
coast of Florida.  \citet{mld1} reconstructed the annual periodic cycle of the mixed layer depth  north to
our target area, numerical values in the coastal region change between 25 and 50 m (deepest in winter).
However, \citet{eddyz1} obtained mean depths of 240 m and 530 m for cyclonic and anticyclonic mesoscale eddies 
in the Peru-Chile Current System exhibiting many similarities to the California Current System.
But the picture is not entirely clear, subsequent high-resolution simulations in the Southern California Bight by 
\citet{eddy12} indicated that most eddies detectable at the surface can reach a depth less than 50 m. The number of 
eddies which penetrate deeper than 50 m decreases dramatically \citep[see Fig.~16 of][]{eddy12}. The mean vertical 
eddy kinetic energy profile steeply drops by the depth, the value at 100 m is around 30 \% compared to the surface
\citep[Fig.~3, ][]{eddy12}. A study of mesoscale eddies in the northwestern subtropical Pacific Ocean by 
\citet{eddy13} did not find such strong differences as \citet{eddyz1}, the trapping depth (above this level the 
rotation speed exceeds propagation speed) was obtained between 120-310 m for cyclonic, and 100-380 m for 
anticyclonic eddies, depending on the geographic region. In a recent study, \citet{eddy17} note that in consistence 
with a rapid decay of the eddy-induced temperature/salinity anomalies and velocity perturbations with depth, the 
eddy fluxes in their study area are surface-intensified and confined mainly to the upper 200 m layer \citep[Fig.~7,][]{eddy17}.

\begin{table}
\caption{Estimates of an effective eddy kinematic viscosity $\nu$ for different wind forcing fractions and integration depths, in units of m$^2$/s.}
\centering
\begin{tabular}{cc|c|c|l}
\cline{3-4}
& & $Z=20$ m & $Z=50$ m  \\ \cline{1-4}
\multicolumn{1}{ |c  }{\multirow{2}{*}{$S_A=$ F$_A$} } &
\multicolumn{1}{ |c| }{shore} &  $0.96 \times 10^{-2}$ &   $0.38 \times 10^{-2}$  &  \\ \cline{2-4}
\multicolumn{1}{ |c  }{}                        &
\multicolumn{1}{ |c| }{open} & $0.44 \times 10^{-2}$ & $0.18 \times 10^{-2}$     \\ \cline{1-4}
\multicolumn{1}{ |c  }{\multirow{2}{*}{$S_A=10\cdot$F$_A$} } &
\multicolumn{1}{ |c| }{shore} & 5.28 $\times 10^{-2}$ & 2.11 $\times 10^{-2}$ \\ \cline{2-4}
\multicolumn{1}{ |c  }{}                        &
\multicolumn{1}{ |c| }{open} & $2.43 \times 10^{-2}$ &   $0.97 \times 10^{-2}$  \\ \cline{1-4}
\end{tabular}
\end{table}

\begin{figure*}
\centering
\noindent\includegraphics[width=0.95\textwidth]{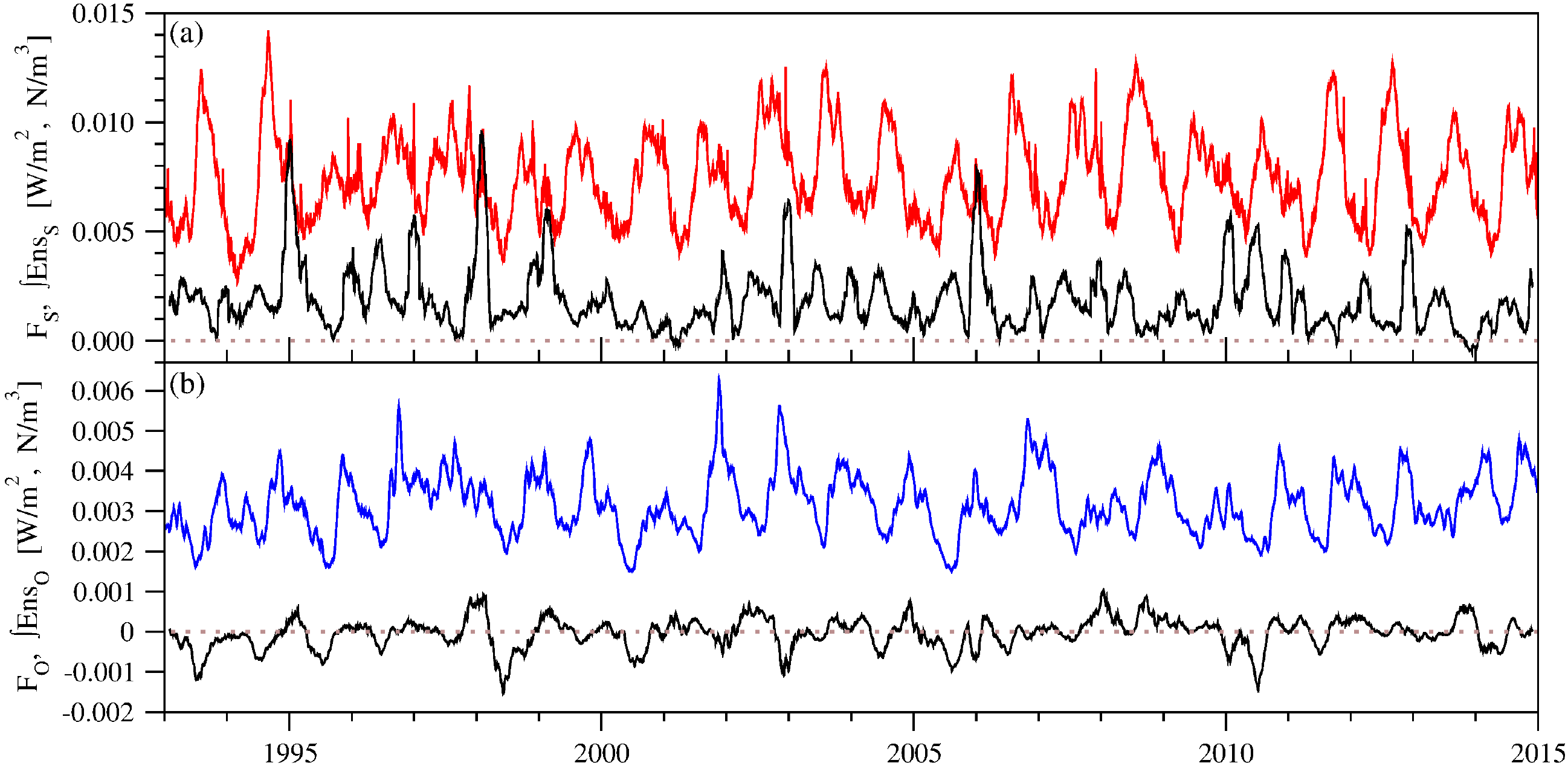}
\caption{Terms of the energy budget  in  Eq.~(7) (see the main text),
(a) shore region and
(b) open water region. Black curves indicate the wind stress work per unit area $\textrm{F}_A$,
61 days running mean is used for smoothing.
Red/blue curves denotes the integrated enstropy per unit area (integration depth $Z=1$ m) for the shore/open
areas.
}
\label{fig:S5}
\end{figure*}

\begin{figure}
\centering
\noindent\includegraphics[width=0.47\textwidth]{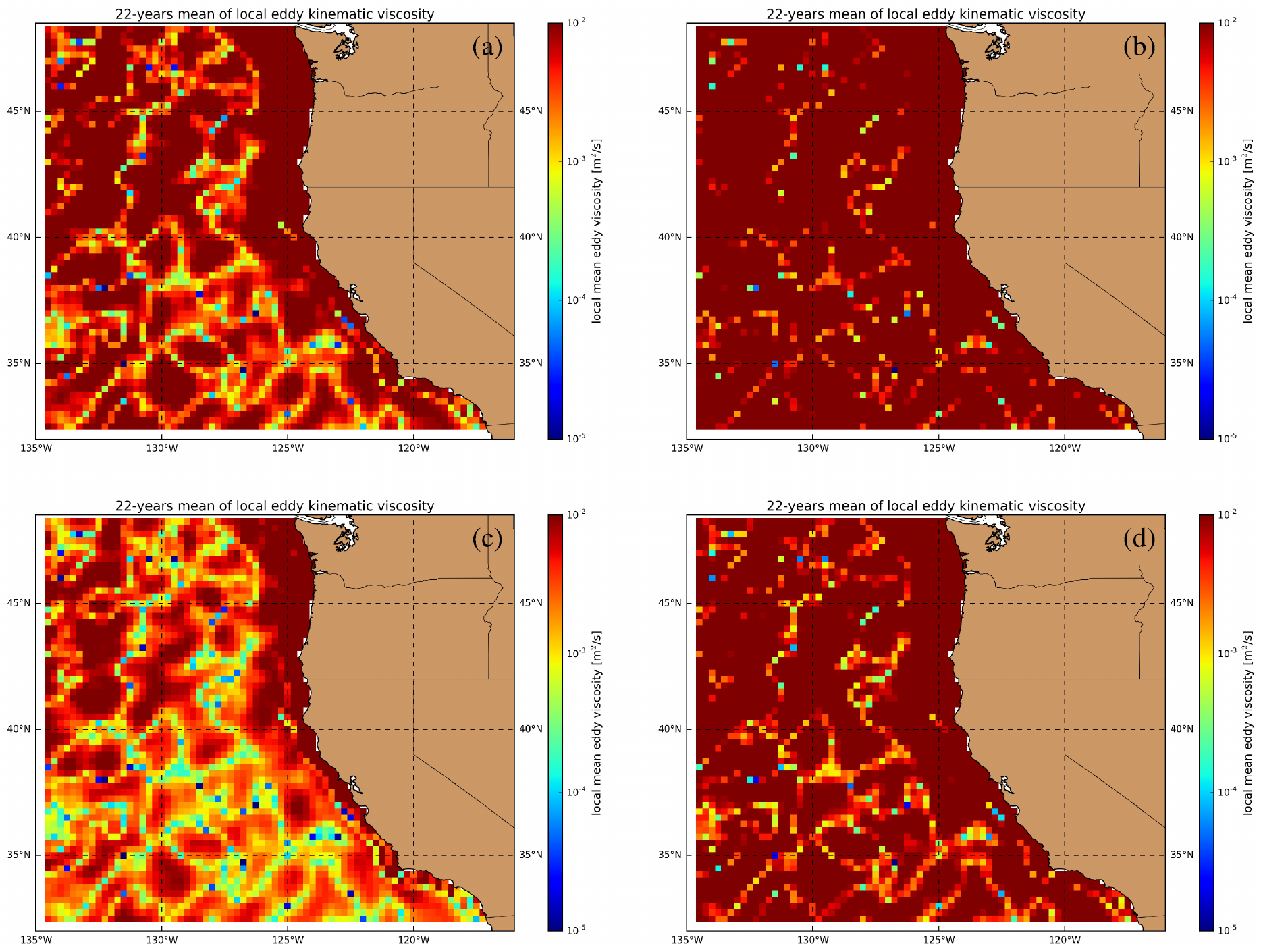}
\caption{ Estimated local eddy viscosity coefficient for each grid cell. The same method is used as in the main text for the
open water and shore regions. The parameters of Eq.~(7) are the following:
 (a) $Z=20$ m, $S_A =$ F$_A$ 
 (b) $Z=20$ m, $S_A =$ 10F$_A$
 (c) $Z=50$ m, $S_A =$ F$_A$
 (d) $Z=50$ m, $S_A =$ 10F$_A$.
}
\label{fig:S6}
\end{figure}

\begin{figure}
\centering
\noindent\includegraphics[width=0.47\textwidth]{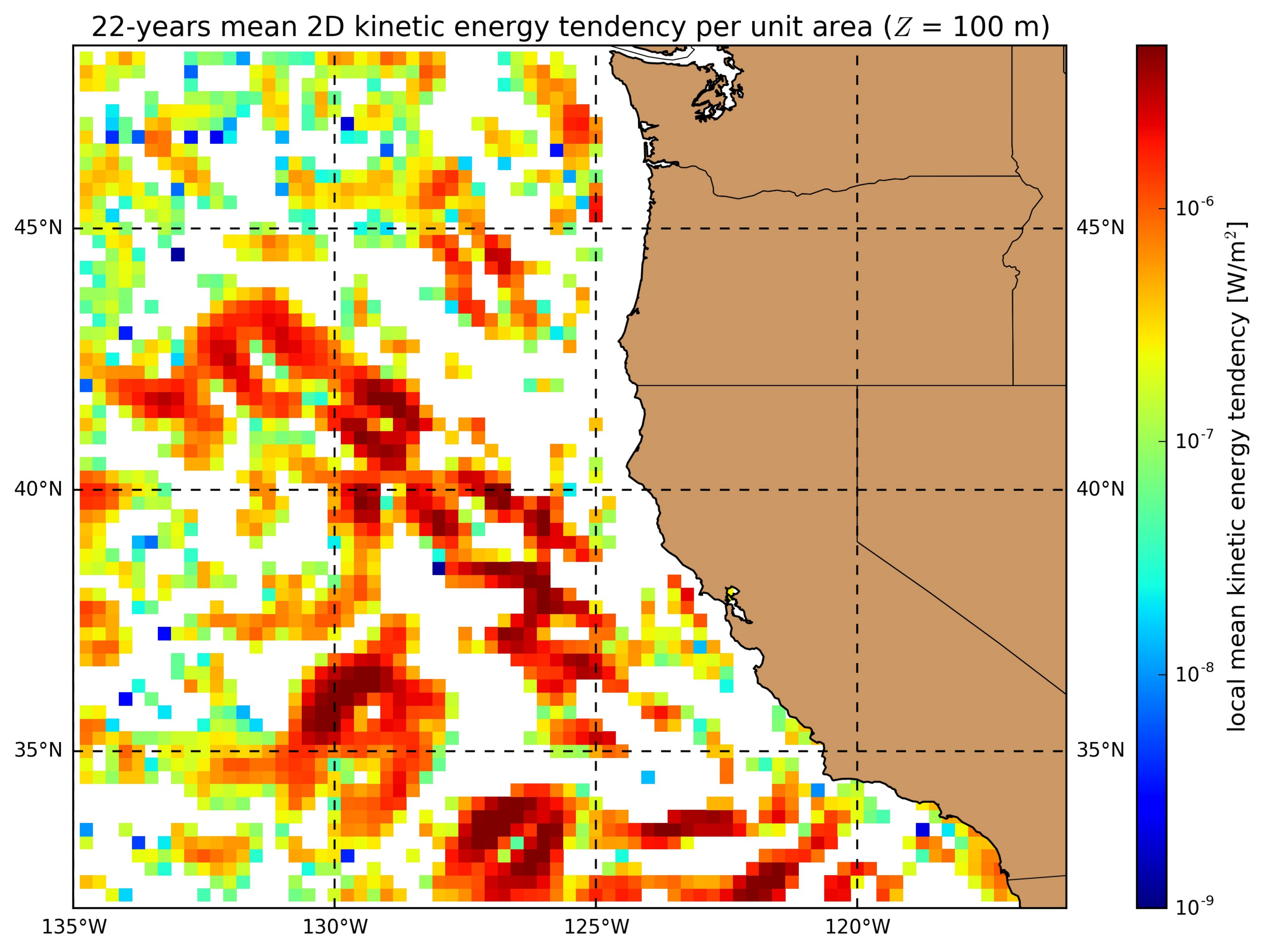}
\caption{ Geographic distribution of the local mean kinetic energy tendency averaged over the 22 years of daily records.
The integration depth for the evaluation was $Z=100$ m [see Eq.~(6) in the main text]. White grid cells belong to small negative
mean values.
}
\label{fig:S7}
\end{figure}

\end{document}